\documentclass[prl,twocolumn,
nofootinbib, preprintnumbers
]{revtex4}

\usepackage{wrapfig}

\usepackage{amsmath,amssymb,amsfonts,times,graphicx}
\usepackage[bookmarks=true,colorlinks,linkcolor=red,urlcolor=blue,citecolor=blue]{hyperref}
\usepackage[usenames]{color}
\usepackage{dcolumn} 

\usepackage[normalem]{ulem}
\usepackage{enumerate}
\usepackage{epsfig}
\usepackage{yfonts}

\usepackage{bm}

\def\greencom#1{#1}
\def\bluecom#1{#1}
\def\redcom#1{#1}

\def\eg{{\it e.g. }}
\def\ie{{\it i.e.}}

\def\({\left(}
\def\){\right)}
\def\[{\left[}
\def\]{\right]}
\def\<{\langle}
\def\>{\rangle}

\def\CA{{\cal A}}
\def\CB{{\cal B}}

\def\CC{{\cal C}}

\def\CF{{\cal F}}

\def\CH{{\cal H}}
\def\CI{{\cal I}}

\def\CT{{\cal T}}

\def\CN{{\cal N}}
\def\CP{{\cal P}}

\def\dslash{\rlap{{\hskip0.2em/}}\partial}

\newcommand\half{{\ensuremath{\frac{1}{2}}}}

\newcommand\ket[1]{\ensuremath{\lvert{#1}\rangle}}

\newcommand{\TT}{\field{T}}

\newcommand{\be}{\begin{equation}}
\newcommand{\ee}{\end{equation}}
\newcommand{\bea}{\begin{eqnarray}}
\newcommand{\eea}{\end{eqnarray}}
\newcommand{\bwt}{\begin{widetext}}
\newcommand{\ewt}{\end{widetext}}

\newcommand{\bi}{\begin{itemize}}
\newcommand{\ei}{\end{itemize}}
\newcommand{\ben}{\begin{enumerate}}
\newcommand{\een}{\end{enumerate}}
\newcommand{\bca}{\begin{cases}}
\newcommand{\eca}{\end{cases}}
\newcommand{\bln}{\begin{align}}
\newcommand{\eln}{\end{align}}
\newcommand{\bst}{\begin{split}}
\newcommand{\est}{\end{split}}

\newcommand{\IP}{{\mathbb P}}

\newcommand{\IR}{{\mathbb R}}
\newcommand{\IZ}{{\mathbb Z}}

\def\Pf{{\rm Pf\ }}

\def\eps{\epsilon}

\def\Ione{\hbox{$1\hskip -1.2pt\vrule depth 0pt height 1.53ex width 0.7pt
                  \vrule depth 0pt height 0.3pt width 0.12em$}}

\def\ii{{\bf i}}

\def\CK{{\cal K}}

\def\SS{{\bf S}}
\def\TT{{\bf T}}

\def\ssigma{{\boldsymbol\sigma}}

\def\dd{\text{d}}

\begin{document}

\title{A gauge theory generalization of the 
fermion-doubling theorem}

\preprint{MIT-CTP/4468, UCSD-PTH-13-07}

\author{S.~M.~Kravec}
\affiliation{
Department of Physics, 
University of California at San Diego,
La Jolla, CA 92093}

\author{John McGreevy\footnote
{On 
leave from: Department of Physics, MIT,
Cambridge, Massachusetts 02139, USA.}}

\affiliation{
Department of Physics, 
University of California at San Diego,
La Jolla, CA 92093} 

%

\begin{abstract}
It is possible to characterize certain states of matter by 
properties of their 
edge states.  
This implies a notion of `surface-only models': 
models which can only 
be regularized at the edge of a higher-dimensional system.
After incorporating the fermion-doubling results of 
Nielsen and Ninomiya into this framework,
we employ this idea to identify
new obstructions to symmetry-preserving regulators of 
quantum field theory.
We focus on an example which
forbids regulated models of 
Maxwell theory with manifest electromagnetic duality symmetry.

\end{abstract}

June 2013

\maketitle

\tableofcontents

\section{Introduction}

This paper is about obstructions to symmetry-preserving regulators
of quantum field theories (QFTs), 
in 3+1 spacetime dimensions.  
The most famous example of such an obstruction is the theorem
of Nielsen and Ninomiya 
\cite{Nielsen:1980rz, Nielsen:1981xu, Nielsen:1981hk,Friedan:1982nk}.
The basic statement of this theorem forbids 
regularization of free fermions 
preserving chiral symmetry\footnote{In odd spacetime dimensions, chiral symmetry is replaced by parity symmetry in this statement.}.
We will approach the study of such obstructions by thinking 
about certain states of matter, 
in one higher dimension, with an energy gap 
(\ie\ the energy of the first excited state is strictly larger than
the energy of the groundstate, 
even in thermodynamic limit).
More precisely, we will study the low-energy effective field theories
of such states; 
below the energy gap, these are topological field theories (TFTs) 
in $4+1$ dimensions.  
Such states will be difficult to realize in the laboratory.
We will use them to demonstrate
an obstruction to any regularization 
of Maxwell theory which preserves manifest
electric-magnetic duality.

To convey the idea of how the study of such extra-dimensional 
models can be useful for understanding the practical question 
of symmetry-preserving regulators of 3+1-dimensional QFTs, 
we must digress on the subject of realizations of symmetries in QFT and in 
condensed matter.  
A basic question in condensed matter physics is
to enumerate the possible gapped phases of matter.
Two gapped phases are equivalent 
if they are adiabatically connected
(varying the parameters in the Hamiltonian whose ground state they are 
to get from one to the other, without closing the energy gap).
An important possible distinguishing feature of different phases is
how the symmetries of the Hamiltonian are realized.
This leads to Landau's criterion which characterizes states
by what symmetries of the Hamiltonian are broken by the groundstate.
Considering this to be understood, it is interesting to refine the question 
to ``How do we distinguish gapped phases that do not break any symmetries?"  

A sophisticated answer to this question, vigorously advocated by Wen 
\cite{wen04, Wen:2012hm},
is {\it topological order}.
A phase with topological order can be characterized by three related properties\footnote{These are sufficient conditions; a 
complete characterization is possible in two dimensions
in terms of adiabatic modular transformations \cite{Wen:2012hm} 
the generalization to higher dimensions seems not to be known yet.}:
\begin{enumerate}
\item {\it Fractionalization of symmetries}, 
that is, emergent excitations
which carry statistics or quantum numbers which 
are fractions of those of the constituents.
\item {\it Topology-dependent groundstate degeneracy}; 
this is a consequence of property 1,
since the groundstates must represent the algebra of flux 
insertion operators associated with adiabatic braiding of 
quasiparticle-antiquasiparticle pairs.
\item {\it Long-range entanglement}, 
which may be quantified 
\cite{LW0605, KP0604}
by the topological entanglement entropy
$\gamma$, a universal piece
of the von Neumann entropy of 
the reduced density of region $A$ of surface area $\ell(A)$
in the groundstate: $ S(A) = \ell(A) \Lambda - \gamma $
(where $\Lambda$ is nonuniversal).
For abelian states, $\gamma$ is 
the logarithm of the number of torus groundstates,
and vanishes for states with short-ranged entangled states
\cite{AshvinTarunTurner}.
\end{enumerate}

Topological order is interesting and difficult.  
Recently, a simpler question has been fruitfully addressed: 
``What are possible (gapped) phases that don't break symmetries and don't have topological order?"
(For a nice review, see the second part of \cite{Turner:2013kp}.)
In this paper we will use the spatial-topology-independence of the 
groundstate degeneracy as our criterion
for short-range entanglement (SRE).  
The $E_8$ state in 2+1 dimensions 
\cite{K0602, Kitaev-unpublished, Lu:2012dt, BGS-unpublished}
is a known exception to this characterization.

A way to characterize such nearly-trivial states is to study them 
on a space with boundary.
A gapped state of matter in $d+1$ dimensions  
with short-range entanglement 
can be (at least partially) characterized 
(within some symmetry class of Hamiltonians)
by {(properties of)} its edge states 
(\ie~what happens at an interface with the vacuum,
 or with another SRE state).
 The idea is simple: 
 if we cannot adiabatically deform the Hamiltonian in time
 from one state to another, 
 we must also not be able to deform the Hamiltonian {\it in space}
 from one state to another,
 without something interesting happening in between.
The SRE assumption is playing an important role here: 
{we are assuming that the bulk state has short-ranged correlations,
so that changes we might make at the surface 
cannot have effects deep in the bulk.}

A useful refinement of this definition incorporates
symmetries of the Hamiltonian:
An {\it symmetry-protected topological (SPT) state}, protected by a symmetry group $G$, is
{a SRE, gapped state, which is not adiabatically connected to a product state
by local Hamiltonians preserving $G$.}
Prominent examples include free fermion topological insulators in 3+1d, protected by $U(1)$ and $\IZ_2^\CT$, 
which have an odd number of Dirac cones on the surface.
Free fermion topological insulators have been classified
\cite{Kitaev:2009mg, Ryu:2010zza}.  Interactions affect the 
connectivity of the phase diagram in both directions: 
there are states which are adiabatically connected only through
interacting Hamiltonians \cite{Fidkowski-Kitaev}, 
and there are states which only exist 
with interactions, 
including all SPT states of bosons \cite{Senthil:2012tm, Vishwanath:2012tq, Wang:2013xqa, Xu:2013bi}.

\begin{wrapfigure}[12]{r}{0.10\textwidth}
  \vspace{-25pt}
  \begin{center}
  \includegraphics[width=0.10\textwidth]{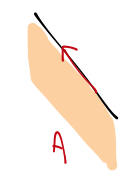}
  \includegraphics[width=0.10\textwidth]{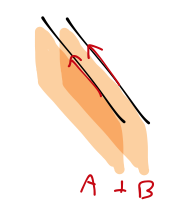}
  \end{center}
\end{wrapfigure}
A simplifying property is that the set of SPT states 
(protected by a given symmetry group $G$) forms a group.
\begin{wrapfigure}[5]{l}{0.2\textwidth}
  \vspace{-25pt}
  \begin{center}
  \includegraphics[width=0.2\textwidth]{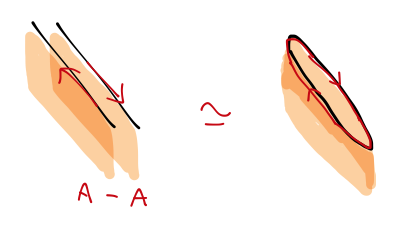}
  \end{center}
\end{wrapfigure}
The addition law is 
tensor product of Hilbert spaces, 
and addition of all interaction terms allowed by $G$.
The inverse operation is simply addition
of the mirror image.
We emphasize that 
with topological order, destroying the edge states
does not modify the nontrivial physics 
(\eg fractional charges) in the bulk:
states with topological order do not form a group.
A conjecture for the 
identity of this group (in $d$ space dimensions, for given $G$)
is the group cohomology group $H^{d+1}(G, U(1))$
\cite{ChenGuWen}.
Exceptions to this conjecture have been found 
\cite{Vishwanath:2012tq,Wang:2013xqa,Metlitski:2013uqa,Burnell:2013bka}
and are understood by
\cite{Kitaev-unpublished}.
For our present purposes, we do not need to know how to classify these states.
The existence of this group struture has the following implication 
\cite{Vishwanath:2012tq,Wang:2013xqa,Metlitski:2013uqa,Burnell:2013bka,Wen:2013oza}, which we may pursue by examples.

Suppose that the edge theory (\eg~with vacuum) of a $(D+1)$-dimensional SPT${}_G$ state
were realized {\it otherwise} -- that is, intrinsically 
in $D$ dimensions, with a local Hamiltonian respecting $G$.  
Then we could paint the conjugate local theory on the surface 
of a $(D+1)$-dimensional space hosting the SPT state 
without changing anything about the bulk.  
By allowed ($G$-preserving, local) deformations of the surface Hamiltonian,
we could then completely destroy the edge states.
But this contradicts the claim that we could characterize the 
$(D+1)$-dimensional SPT state by its edge theory.
We conclude 
that 
the edge theories of SPT${}_G$ states cannot be regularized intrinsically in $D$ dims, 
while preserving $G$.  We will refer to them as ``surface-only models".
More generally, we can consider the interface between {\it pairs}
of SRE states;
the edge theory for interface between any pair of SPT states
is also a surface-only theory.

The Nielsen-Ninomiya result is implied by this logic.
Consider free massive (relativistic, for convenience) fermions 
in 4+1 dimensions: $ S = \int d^{4+1}x \bar \Psi\( \dslash + m \)  \Psi$.  
The two signs of $m$ label distinct phases.
One proof of this arises by coupling to an external gauge field 
via the fermion number current
$ \Delta S= \int d^5x A^\mu \bar \Psi \gamma_\mu \Psi $. 
The effective action 
involves a quantized topological term: 
$$ \log \int [D\Psi] e^{ \ii S_{4+1}[\Psi, A]} 
=
{ m \over |m| }
\int {\dd^5 x   \over 24 \pi^2}  \epsilon_{abcde} A_a F_{bc} F_{de}. $$
The domain wall between the two phases
hosts an exponentially localized chiral fermion field 
\cite{Callan:1984sa, Kaplan:1992bt},
a fact which has been useful for lattice simulations \cite{Kaplan:1992bt,Kaplan:2009yg}.

A more famous $D=3+1$ analog begins with a massive fermion in 3+1 dimensions:
$$ S = \int d^{3+1}x \bar \Psi\( \dslash + m + 
\ii \hat m \gamma^5 \)  \Psi ~~$$ 
If we demand time-reversal ($\CT$) invariance, $\hat m=0$,
and $\pm m$ label distinct SPT states protected by $\IZ_2^\CT$.
Coupling to an external gauge field
$$ \log \int [D\Psi] e^{ \ii S_{3+1}[\Psi, A]} 
= { m \over |m| } \int { \dd^4 x \over 32 \pi^2} \epsilon^{abcd} F_{ab} F_{cd}.$$
produces a quantized magnetoelectric effect with $\theta = 0, \pi$
\cite{Qi:2008ew}.
The domain wall hosts a single Dirac cone in 2+1d.
(We must emphasize that in both the above examples
the protecting symmetry which distinguishes 
these states from the trivial insulator
includes charge conservation.)

This suggests the following strategy for
generalizing these results away from free fermions.
Study a simple \greencom{(unitary)} gapped or topological field theory in 4+1 dimensions 
\redcom{without topological order}, \greencom{with symmetry $G$}.
Consider the model on the disk
\greencom{with some boundary conditions.}
The resulting edge theory 
is 
a ``surface-only theory with respect to $G$'' 
\bluecom{-- it cannot be regulated by a local 
$3+1$-dim'l model while preserving $G$.}

What does it mean to be a surface-only state?
Such a model is perfectly consistent and unitary; it can be realized at the 
edge of some gapped bulk theory.   However, it 
cannot be regularized in a local way consistent with the symmetries in absence of the bulk.

It {(probably)} means these QFTs will not be found 
as low-energy EFTs of solids or in cold atom lattice simulations.
Why `{probably}'?  This perspective does not rule out emergent (``accidental") symmetries,
not explicitly preserved in the UV. 
An example of a SPT-forbidden symmetry emerging in the IR occurs 
in critical Heisenberg spin chains, where the 
spin symmetry is enhanced to an O$(4)$ which 
rotates the spin order into the valence bond solid order.

It also does not rule out symmetric UV completions that include gravity,
or decoupling limits of gravity/string theory.
\redcom{(UV completions of gravity have their own complications!)} 
\bluecom{String theory strongly suggests the existence of Lorentz-invariant states of gravity 
with chiral fermions and lots of supersymmetry} 
\greencom{(the $E_8\times E_8$ heterotic string, 
chiral matter on D-brane intersections, self-dual tensor fields...)}
some of which can be decoupled from gravity.

\section{A simple family of topological field theories in 4+1 dimensions}

Here we will study 
simple field theories in 4+1 dimensions,
whose path integral is gaussian.
We follow an analog of the 
K-matrix approach used in
\cite{Lu:2012dt} to study 2+1 dimensional SPT states.

\def\nB{N_B}
Specifically, consider $2\nB$ $2$-form potentials $B^I_{MN}$,  in 
$4+1 $ dimensions, with action
\be \label{eq:CS} S_\text{CS}[B] = {K_{IJ}\over 2 \pi } \int_{\IR \times \Sigma_{4}}  B^I \wedge \dd B^J ~.\ee
$M,N = 0..4$ is a 4+1 dimensional spacetime index, $I,J = 1..2\nB$, 
and 
$\Sigma$ is a smooth 4-manifold (sometimes with boundary).

In $4\ell +1 $ dims, $K$ is a {\it skew}-symmetric integer $2\nB \times 2\nB$ matrix.
This basic difference from 2+1 dimensional CS theory
(where the K-matrix is symmetric) arises 
from the fact that the wedge product of two-forms is symmetric, 
and hence $ B \wedge \dd B  = \half \dd (B \wedge B) $,
and thus the symmetric part of $K$ produces a total derivative.
The action is independent of a choice of metric on $\IR \times \Sigma_{2p}$.
Just as one may add an irrelevant Maxwell term to CS gauge theory, 
these models may be considered as the $g\to \infty$ limit of 
the non-topological models with (`topologically massive' \cite{Deser:1981wh}) propagating two-forms with action
\be \label{eq:kinetic} 
S = \int {\tau_{IJ}\over M}  \dd B^I \wedge \star \dd B^J + S_\text{CS}[B] ~;\ee
the mass scale $M$ determines the energy gap.

These models have a long history, including
\cite{Schwarz:1978cn, Schwarz:1979ae, Witten:1988hf, Elitzur:1989nr, Horowitz:1989ng, Blau:1989bq, Horowitz:1989km, Witten:1998wy, 
Moore:2004jv,
Belov:2004ht, Hartnoll:2006zb}.
The canonical quantization of closely-related models was discussed in 
\cite{Horowitz:1989ng, Blau:1989bq}.
(The difference from the models we study comes from issues related to the compactness of the gauge group,
as in \eqref{eq:largegauge} below.
This includes the quantization of the level and 
the resulting conclusion about the finiteness of the  dimension of the Hilbert space.)
The linking number of (homologically-trivial) pairs of surfaces was constructed
as an observable of this theory in
\cite{Horowitz:1989km, Blau:1989bq}.

Moore \cite{Mooretalk} refers to these models as both trivial and difficult.
They are trivial in that the path integral is gaussian,
and difficult in the sense that there are lots of subtleties in defining 
what's being integrated over.  
We will not highlight all these subtleties, but we will endeavor to not be wrong\footnote{This paper \cite{Freed:2006yc} gives a readable description 
of the dangers here.  We can go a long way toward avoiding running afoul
of the subtleties which are resolved by 
formulating the model in terms of differential cohomology
if we only consider 
manifolds whose homology has no torsion.  Even with this restriction
we can say something interesting.  More subtle distinctions between 4+1d phases
that only arise on manifolds with torsion in their homology will have to wait for future work.}.
Moore and collaborators \cite{Mooretalk}
have studied such generalizations of CS theory
and their edge states with great care.
In particular, the paper
\cite{Belov:2004ht}
constructs the partition function for the theory we study below
and relates it to a piece of the $\CN=4$ super Yang-Mills partition function.

This relation arises because these models are realized as a `topological sector' 
of type IIB strings on $AdS_5 \times X_5$ \cite{Witten:1998wy};
the simplest case with two 2-forms arises for $X_5 = S^5$,
where the forms may be identified as 
$B^1\equiv B_{NSNS}, B^2\equiv C_{RR}$
which couple to fundamental strings and D1-branes respectively; 
the CS term 
\eqref{eq:CS} arises from 
$$ S_{IIB} \ni 
{1\over 2\pi} \int_{AdS_5\times S^5} F_{RR}^{(5)} \wedge B \wedge \dd C
= {N \over 2\pi}  \int_{\IR \times \Sigma}  B \wedge \dd C $$
and allows for 
the ending of $N$ F-strings 
holographically dual to the baryon vertex 
of $\CN=4$ SYM
\cite{Gross:1998gk, Witten:1998xy}.
$S$-duality of IIB string theory 
acts by interchanging the two-forms.
This symmetry of \eqref{eq:CS} will play a crucial role below.

We focus on the Abelian case.  
The action $S$ is invariant under the gauge redundancies
\be B^I \simeq B^I + \dd \lambda^I \ee
where $\lambda^I$ are one-forms.  
We will impose large gauge equivalences:
\be\label{eq:largegauge} B^I \simeq B^I + n^\alpha \omega_\alpha , ~~~
[\omega^\alpha] \in H^2(\Sigma, \IZ), ~~ n^\alpha \in \IZ^{b^2(\Sigma)} \ee
where $ b^2 \equiv \text{dim}H^2(\Sigma, \IZ)$ is the 2nd Betti number of $\Sigma$.
In the case of 2+1 CS gauge theory of $\IR \times \Sigma_2$, the
analogous identification
arises naturally from large gauge transformations \cite{Witten:1988hf, Elitzur:1989nr}
$$ A \equiv A + \ii g^{-1} \dd g,
~~\text{with}~~~g(x) = e^{ \ii \sum_{\alpha=1}^{b^1(\Sigma_2)}\int_{x_0}^x n^\alpha \omega_\alpha} $$
where $ \omega_\alpha$ form a basis of $H^1(\Sigma_2, \IZ)$, and $x_0$ is a base point.  
In the 4+1 dimensional case, we don't know what a group-valued one-form is, but retain the
natural identification \eqref{eq:largegauge}.
(A mathematical formalism which produces this identification 
in a `natural' way is described in 
\cite{Freed:2006yc}.)
This identification was not imposed in the otherwise-identical models 
studied in \cite{Horowitz:1989ng}.

This class of models has been used \cite{Shatashvili-unpublished, 
Witten:1996md,
Maldacena:2001ss, Moore:2004jv,
Belov:2004ht,Belov:2005ze,Belov:2006jd,Mooretalk}
to `holographically' {\it define} the partition function of the edge theory.
(These papers focus mainly on the case
of bulk spacetime dimension $D=4\ell +3$: 1+1d chiral CFTs, conformal blocks of the 5+1d (2,0) theory.)
In this paper, we are using this same relation to a different end.

Finally, we note that the simplest model 
\eqref{eq:CS} with one pair of two-forms is equivalent to a $\IZ_k$ {\it 2-form} gauge theory
\cite{Maldacena:2001ss, 2004AnPhy.313..497H}.  
The case we will be most interested in, with $k=1$, 
can therefore be regarded as a kind of `$\IZ_1$ gauge theory',
which nevertheless has a something to teach us.

\section{Bulk physics}
\subsection{When is this an EFT for an SPT state?}

For this subsection we suppose that $\partial \Sigma$ is empty.
We wish to ascertain the size of the Hilbert space,
and its dependence on the topology of $\Sigma$.
Thinking of this as the EFT for some 4+1 dimensional gapped state of matter,
at energies below the gap, this
number is the groundstate degeneracy
(up to $e^{-\text{(system size)} \cdot \text{gap} } $ finite-size effects).  
If this degeneracy is dependent on the topology of $\Sigma$,
then this state has topological order in the bulk \cite{wen04}.
A closely related calculation appears in section 6 of 
\cite{Freed:2006yc}, in the case of a theory of $p$-forms in $2p+1$ dimensions,
for the case of odd $p$.

Many aspects of the problem are analogous to the case of 2+1 dimensional 
CS 1-form gauge theory.   
The kinetic term in \eqref{eq:kinetic} is analogous to the Maxwell term,
and $g$ has units of energy.
With $g<\infty$ in \eqref{eq:kinetic}, 
there are excited states 
of energy $ E \propto g$, 
analogous to higher Landau levels.
As in 2+1d, this is an example of an inclusion of metric dependence which 
does not change the nature of the phase.
We will focus on the limit $g \to \infty$.

And as in the 2+1d case, the identification \eqref{eq:largegauge} means that 
the gauge-inequivalent operators 
are labelled by cohomology classes, here $[\omega_\alpha] \in H^2(\Sigma, \IZ)$: 
\def\xxi{m}
$$ \CF_\omega(\xxi) \equiv e^{ 2 \pi \ii \xxi_I \int_\omega B^I} $$
where the identification on $B$ implies 
$ \xxi^I \in  \IZ$.  
These operators satisfy a Heisenberg algebra:
$$ \CF_{\omega_\alpha}(\xxi)
\CF_{\omega_\beta}(\xxi')
= \CF_{\omega_\beta}(\xxi')
\CF_{\omega_\alpha}(\xxi) e^{ 2 \pi \ii  \xxi_I^\alpha \xxi_J^{'\beta} {\(K^{-1}\)^{IJ} \CI_{\alpha\beta} }}. $$
This kind of algebra is familiar from the quantum Hall effect,
and from 2+1 dimensional CS theory.
Its only irreducible representation is 
the Hilbert space of a particle on a `fuzzy torus.'
of dimension $2 \nB b^2(\Sigma)$.  
The number of states depends on the intersection form on the second homology of $\Sigma$:
$$ \int_{\Sigma} \omega_\alpha \wedge \omega_\beta = \CI_{\alpha\beta}. $$
This is a $b^2(\Sigma)\times b^2(\Sigma)$ {\it symmetric} matrix, which 
has various properties guaranteed by the theory of 4-manifold topology \cite{DonaldsonBook}.  
For a smooth, compact, oriented, simply-connected 4-manifold,
it is unimodular (\ie\ has $|\det \CI | = 1$).
It is even if $\Sigma$ is a spin manifold
(\ie\ it admits spinor fields, \ie\ its first Steiffel-Whitney class vanishes).
Its signature is not definite.

Consider the simplest nontrivial case where $b^2(\Sigma)=1$,
which case $\CI$ is a $1\times 1$ matrix.
The simplest example is $\Sigma = \IP^2$ where $ \CI = 1$ \cite{DonaldsonBook}.

We may arrive at the same conclusion by expanding the action in a basis of cohomology 
representatives: $$ B^I = \sum_{\alpha=1}^{b^2(\Sigma_4, \IZ)}  \omega_\alpha b^{I\alpha}(t) , ~~~
\text{span}\{ \omega_\alpha \} = H^2(\Sigma_4, \IZ) , $$
so
$$ S = { K_{IJ} \over 2 \pi} \int dt \int _{\Sigma_4 } \omega_\alpha \wedge \omega_\beta 
b^{I\alpha} \dot b^{J\beta} 
= { K_{IJ} \over 2 \pi} \int dt \CI_{\alpha \beta} b^{I\alpha} \dot b^{J\beta} 
$$
which describes a particle in $ b^2(\Sigma) $ dimensions with a magnetic field in 
each pair of dimensions of strength $k$.

First we further assume that there is just one pair of $B$-fields, 
$\nB=1$ and 
(WLOG by a $GL(\nB, \IZ)$ rotation) we take
$K = k \ii \ssigma^y$.  Then the hilbert space is that of a particle on a periodic 1d lattice with $k$ sites:
$$ \CF_1 \ket{\Omega}= \ket{\Omega}, ~ \ket{w} = \CF_2^w \ket{\Omega}, ~~ w = 1... k .$$
Here $\CF_I \equiv \CF_I(1)$.
(Note that $ \CF_I^k = \Ione$.)

Now keep this simplest $\Sigma$ but take more pairs of $B$-fields.
We can skew-diagonalize the $K$-matrix, so that 
we find $\nB$ copies of the previous discussion, with various $k$,
which are the skew-eigenvalues of $K$: 
$$ K = \begin{pmatrix}   0 & k_1  &  0 & 0 & \dots  \\
- k_1 & 0 & 0 & 0 & \dots \\ 
0 & 0  &  0 & k_2 & \dots  \\
0 & 0 & -k_2 & 0 & \dots  \\
\vdots & \vdots & \vdots & \vdots & \ddots  
\end{pmatrix}  ~.$$
So the dimension of the Hilbert space of the bulk in this case is 
$$ \dim \CH(b_2(\Sigma)=1, \CI = 1) = \prod_{a = 1}^{\nB} k_a = \Pf K,$$
the pfaffian.

If $\CI > 1$, it multiples $k_a$ in the commutation relation.  This gives 
$$ \dim \CH(b_2(\Sigma)=1, \CI ) = \prod_{a = 1}^{\nB} k_a = \Pf \( \CI K\) .$$

Now consider a more general 4-manifold where $b_2 > 1$.
Then there are $ b_2 \nB$ pairs of canonically-conjugate variables.
By $ GL(2 b_2 \nB, \IZ)$ rotations, we can choose a basis to diagonalize 
$ \CI \otimes \CK$.  
Since $\CI$ is symmetric, this matrix is still skew-symmetric.
If we call its skew-eigenvalues $ \lambda_A$,
and the number of groundstates is 
$$ \boxed{\dim \CH = \prod_{A = 1}^{b_2(\Sigma) \nB} \lambda_A = \Pf \( \CI \otimes \CK \)}~.$$

The conclusion from this discussion is that if $\CI \otimes K$ 
has any skew-eigenvalues which are larger than one, 
the system has bulk topological order: 
a topology-dependent groundstate degeneracy.

\subsection{This is a model of bosons}

We observe that if $\Sigma$ is a spin manifold,
then we may take the entries in $K$ to be half-integer,
while preserving the consistency of the theory
(in particular, the fact that the number of groundstates is an integer).
A choice of spin structure was not required for the calculation we did here, 
but would be for other observables. 
Such models would provide low-energy effective theories
for fermionic models, analogous to 
odd $k$ CS gauge theories in two dimensions,
as is familiar from the integer quantum Hall effect
and was studied in great detail by 
\cite{Belov:2005ze}.

The preceding statements provide low-energy evidence that 
the models \eqref{eq:CS} for integer $K$
can arise as the low-energy EFT for 
models of bosons.  
High-energy evidence for this claim comes from
\cite{BGSprogress},
which constructs a local bosonic lattice model
which produces this EFT at low energies.

\section{Surface states}

\subsection{2+1d case with boundary}

Here we briefly review some relevant properties of 2+1d CS gauge theory
on a spacetime of the form $ \IR\times \Sigma$ with 
a spatial boundary, $ \partial \Sigma$
\cite{Witten:1988hf, Elitzur:1989nr, Maldacena:2001ss, Belov:2005ze, wen04}.
We will further assume for convenience that $H^p(\Sigma, \partial \Sigma) = \delta^{0,p}$ --
the topology (specifically, the relative cohomology) of the bulk is trivial.

The gauge group consists only of gauge transformations which 
go to the identity at the boundary. 
Transformations which act nontrivially do not leave the CS action invariant;
if we wished to mod out by them we would need to add degrees of freedom to cancel this variation,
and would arrive at the same description.
The CS equation of motion for $A_0$ is a constraint which imposes $ 0 = F$ 
which means that 
\be\label{eq:Agauge} A = g^{-1} \dd g \ee
 which we can choose to vanish in the bulk.
In the abelian case we can take $ A = \dd \phi$, with $\phi \simeq \phi+ 2\pi$ compact.
We arrive at a theory of compact free bosons.

The Hilbert space of edge states is specified universally by the bulk -- the CS action evaluated on 
\eqref{eq:Agauge} gives 
$ S_0 = \int_{\IR \times \partial \Sigma} { k^{IJ} \over 4 \pi} \partial_t \phi^I \partial _x \phi^J $
which determines the canonical equal-time commuators
$$ [ \phi^I (x), \phi^J(x') ] = 2 \pi \ii k^{-1}_{IJ} \theta(x-x') . $$
However, the Hamiltonian governing these bosons comes entirely from the choice
of boundary conditions, 
since the bulk action is linear in time derivatives.  
The non-universal velocity of the chiral edge modes is not encoded in the Chern-Simons action;
this is natural from the quantum Hall point of view,
where this velocity is determined by the slope of the potential confining the 
Hall droplet.
The non-universal velocity term 
arises if we choose 
the `boosted' boundary condition $ A_0 + v A_x |_{\partial\Sigma}= 0 $, 
where $x$ is the coordinate along the boundary.
(Note that this is a boundary condition, not a gauge choice -- $v$ affects the physics.)
This gives 
$ S[\phi] = { k \over 4\pi} \int \( \partial_t + v \partial_x \) \phi \partial_x \phi $,
which shows that the edge boson is chiral; 
the sign of $v$ is correlated with the sign of $k$ in order that the energy be bounded below.
Alternatively, and more covariantly, we may
use units with $v=1$ and impose $ A - \star A |_{\partial \Sigma} =0 $
\cite{Mooretalk}.  
We may neglect the possibility of a matrix of velocities
if we include a coupling to disorder \cite{Kane:1994vb,Kane:1994jd},
which we should unless we are interested in SPT
states protected by translation invariance.

\subsection{4+1d abelian two-form gauge theory with boundary}

As in the previous discussion, we consider 
the bulk spacetime to be $ \IR \times \Sigma_4$ 
with nontrivial boundary $ \partial \Sigma_4$,
and trivial topology in the bulk.
The gauge transformations must act trivially at the boundary.

We will focus on the simplest case where $K = k \ii \ssigma^2$.  Consider 
\bea S[B,C] &=& { k \over 2 \pi} \int_{\IR \times \Sigma_4} 
\( B \wedge \dd C  - C \wedge \dd B \)
\cr 
&+& \int_{\IR \times \partial \Sigma_4} \(  { 1\over 4 g^2 }C \wedge \star_4 C + { k \over 4 \pi} B \wedge C\)
\nonumber
\eea

A convenient boundary condition
(imposed by free variation of \eqref{eq:bcBC} 
with the boundary terms indicated)
is: 
$$\( { k\over 2 \pi } B - { 1\over 2 g^2 } \star_4 C \)|_{\partial \Sigma_4} = 0~. $$
The path integral over $B$ produces a delta function forcing $C$ to be flat \cite{Maldacena:2001ss}: 
$$ \int[DB] e^{ \ii S } = \delta [ \dd C] ~~~\implies ~~~ C = \dd a ~.$$
$$ S[C= \dd a] =  
\boxed{ { 1\over 4 g^2 } \int _{\IR \times \partial \Sigma_4}  \dd a \wedge \star_4 \dd a  }$$

This is ordinary Maxwell theory.
We know how to regularize this theory with a local 
lattice boson model
\footnote
{For example,
\bea
 {\bf H} &=& - \sum_{\text{vertices},v \in \Delta_0} \( \sum_{\ell \in s(v) } n_\ell - q_v \)^2 \cr \cr &-& \sum_{ p \in \Delta_2}  \prod_{\ell \in \partial (p) } e^{ \ii b_\ell} + h.c. 
- \Gamma \sum_{\ell \in \Delta_1}  n_\ell^2 .
\label{eq:bcBC}
\eea
$ \Delta_p \equiv \{ p\text{-simplices} \} $.
$s(v) \equiv \{ \text{edges incident on $v$ (oriented ingoing)} \}$
and $ [b_p, n_p] = \ii $ 
is a number-phase representation; $ b_p \equiv b_p + 2\pi, n_p \in \IZ$.}
We infer from the logic described above
that we are forced to break 
some symmetry of \eqref{eq:CS} in order
to realize its edge states in a local way.
What symmetry must we break in writing such a local model?

We answer this burning question in the next section,
after which we comment on more general boundary conditions.
The general abelian case
produces a collection of copies of Maxwell theory.
This can be seen by block-diagonalizing $K$
by a GL$(2\nB, \IZ)$ rotation.

\section{Symmetries}

CS gauge theories in 2+1d have bulk `topological' conserved currents.
When these theories arise from quantum Hall states, 
some linear combination of these currents
can be interpreted as electron number: 
$$ 0 = \partial^\mu J_\mu 
~~~\implies~~~ J_\mu \equiv {1\over 2\pi} \epsilon_{\mu\nu\rho} \partial_\nu A_\rho ~.$$
$J$ is conserved when $A$ is single-valued. 
In 4+1d, the analog is pairs of string currents
$$ J^I_{\mu\nu} \equiv {1\over 2\pi} \epsilon_{\mu\nu\rho\sigma\lambda} \partial_\rho B^I_{\sigma \lambda} ~.$$ 
We can use these 
symmetries to
demonstrate that different $K$ label nontrivial, different states.

In $D=2+1$ CS theory, we can couple the particle currents $J^I \equiv  \star \dd A^I $ 
to {\it external} 1-form potentials, $ \CA_I$: 
$$ \log \int [DA^I] e^{ \ii  \int {k\over 4 \pi} A\dd A + \ii  \int J_I \CA^I}
=  \int_{2+1}  \(4 \pi k^{-1}\)_{IJ}  \CA^I \dd \CA^J ~~
$$
This term in the effective action demonstrates a quantized Hall response;
in the absence of topological order ($\det k = 1$), 
$\sigma_{xy} {h \over e^2} $ is an integer.
This is one way to distinguish the integer quantum Hall state from a trivial insulator.
In the 2+1d case, the thermal Hall response $ \propto c_L-c_R$ 
makes this distinction even in the absence of charge conservation.

The analogous stratagem in $D=4+1$ is to couple the string currents $J^I =  \star \dd B^I $ 
to {\it external} 2-form potentials, $ \CB_I$: 
$$ \log \int [DB^I] e^{ \ii \int {K\over 4\pi} B\dd B + \ii  \int J_I \CB^I}
=  \int  \(4 \pi K^{-1}\)_{IJ}  \CB^I \dd \CB^J ~~.
$$
This term exhibits a quantized `string Hall' response
to external 3-form field strengths, 
which again is an integer in the absence of topological order, $\text{Pfaff}K = 1$).
This response distinguishes this state from the completely trivial 
state of bosons in 4+1 dimensions.

\begin{itemize}
\item Translation invariance is a red herring. 
In fact, breaking translation invariance with static disorder
helps to produce a uniform speed of light in the edge theory.
Indeed, the lattice model \cite{BGSprogress}
should have the same edge states.

\item Stringy symmetries: 
$\displaystyle{J^B_{\ell 0}|_{bdy} =  E_\ell, 
 J^C_{\ell 0}|_{bdy} = -  B_\ell}. $
\greencom{$ E_\ell \equiv \partial_t a_\ell - \partial_\ell a_t $ 
$ B_\ell \equiv \eps_{\ell ij} (\partial_i a_j  - \partial_j a_i )$ are ordinary E$\&$M fields}
$$ J^C_{y0}  = \epsilon_{ijk} \partial_i C_{jk} = \epsilon_{ijk} \partial_i \partial_j a_k = \vec \nabla \cdot \vec B $$ 
$$ J^B_{y0}  = \epsilon_{ijk} \partial_i B_{jk} = 
\epsilon_{ijk} \partial_i \epsilon_{jkl} E_\ell  = 
\vec  \nabla \cdot \vec E . 
$$
\redcom{This is ordinary charge, of course it has to be conserved.  }
 \item $\CC$: $(B,C) \to - (B,C)$ 
is $ (\vec E, \vec B) \to - (\vec E, \vec B)$.  
\bluecom{This is preserved in pure U(1) lattice gauge theory.}
 \item $\CT \CP$: 
$  t \to - t, x^M \to - x^M, \ii \to - \ii $, 
$ B \to - B , C \to C$ as two-forms. 
\bluecom{Acts in the usual way on the EM field as $ (E, B) \to (E, -B)$.}

\item{} 
$\displaystyle{\boxed{\text{ {\bf Electric-magnetic Duality:} $ (B,C) \to (C, - B)$ }}}$
The global symmetry which interchanges $B$ and $C$
is a manifest symmetry of the bulk theory.  
It acts on the boundary gauge field
as electromagnetic duality $ (\vec E, \vec B) \to (\vec B, - \vec E)$.

\item{} 
Just as in string theory and 
gauge theory \cite{Schwarz:1993vs}, the $\IZ_2$ EM transformation just described 
is a subgroup of a classical SL$(2, \IR)$ symmetry acting 
on $(B,C)^T$ as a doublet $ \begin{pmatrix} B \cr C \end{pmatrix}
\to\begin{pmatrix} a & b \cr c & d \end{pmatrix}
 \begin{pmatrix} B \cr C \end{pmatrix}
 $ with $ ad -bc =1$.
 The continuous parts of this group, where $ a,b,c, d \notin \IZ$
do not preserve the integrality condition \eqref{eq:largegauge},
and are not symmetries of the quantum theory.
The SL$(2,\IZ)$ subgroup is generated by 
$$ \SS = \begin{pmatrix} 0 & 1 \cr -1 & 0 \end{pmatrix} ~\text{  and }~~ 
\TT = \begin{pmatrix} 1 & 1 \cr 1 & 0 \end{pmatrix} . $$
$\SS$ is the EM duality transformation above, and 
$\TT$ adds a boundary term $\int_{\IR \times \partial \Sigma} { k \over 4 \pi}  C \wedge C$;
this shifts the $\theta$ angle of the surface gauge theory by $ 2\pi$.

\end{itemize} 

This EM duality symmetry is effectively unbreakable in the 4+1d CS theory.
Even with arbitrary boundary conditions, 
breaking Lorentz invariance,
the scaling freedom in the relationship between $B,C$ and 
$a, \tilde a$, and in the duality map itself
always allow a symmetry operation exchanging
$ a$ and $\tilde a$.
The general boundary condition is of the form
$$ 0 = \( B + c_1 C + c_2 \star B + c_3 \star C \) |_{\partial \Sigma}~. $$
Of the three real coefficients $c_i$, 
one may be absorbed in the speed of light, 
a second may be absorbed in the relationship between 
$C$ and the Maxwell potential $a$, 
and the third may be absorbed in the duality map, 
$ (B, C) \to ( \lambda C, \lambda^{-1} B)$
\footnote{This transformation also acts on the large gauge identification map \eqref{eq:largegauge}, 
and on the surface acts as the usual (fake) rescaling of the charge lattice by 
$ (e q_e, g q_m) \to (\lambda e q_e, \lambda^{-1} g q_m)$ preserving Dirac quantization $ g e = 2\pi$.}.

The only way to break the manifest EM duality symmetry
of the bulk is to add charged matter,
in the form of strings which end at the boundary.
Any matter we add has a definite
charge vector $(q_e, q_m)$.
Even if the matter comes in dual pairs, 
condensing matter with a definite charge vector
will gap out the boundary photon
(and the matter with the dual charge vector).

Charged matter in the surface Maxwell theory arises
from strings which are minimally coupled to the bulk 2-forms
and terminate at the boundary.
The precise spectrum of matter is information additional
to the low-energy bulk action we have described;
it is specified in a lattice UV completion \cite{BGSprogress}.

We conclude that it is not possible to regularize Maxwell theory in a way
which retains manifest electromagnetic duality symmetry.
This is consistent with the existence of U$(1)$ lattice gauge theory, 
and its UV completions in terms of lattice
models with local Hilbert spaces 
\cite{LW0510, K0602}, in which EM duality is not manifest on the lattice.
Of course, we still do expect this symmetry to emerge in the IR limit
of these models.

\section{Discussion}

\subsection{7d CS theory and the (2,0) superconformal theory}

There are many generalizations of the kind of nearly-trivial
bulk theory we have studied above.
Consider the following 6+1d Chern-Simons theory
$$S_7[C^{(3)}] = { k \over 4 \pi} \int_{\IR \times \Sigma_6} C^{(3)} \wedge \dd C^{(3)} $$ 
For $k=1$, there is no topological order,
and the model is completely trivial in the bulk.
To study the edge states, 
we examine solutions of the bulk equations of motion $ C^{(3)} = d c^{(2)} $, 
with 
the boundary condition: $ C_{0ij} = v (\star_{6} C)_{ij} $.  The resulting action is
$$ S_7[C^{(3)} = d c^{(2)}] = { k \over 4 \pi}  
\int_{ \IR \times \partial \Sigma_6}  \epsilon \partial c^{(2)} \cdot \( \partial_t c^{(2)} + v \epsilon \partial c^{(2)} \)  . $$
$c$ is a self-dual 2-form potential in 5+1d.
In this case (as in the 2+1d CS theory and the integer QHE),
no symmetry is required to protect the edge states.  
We conclude that this model cannot be regularized intrinsically in 5+1 dimensions.

This model describes the `topological sector' of the {\it $A_0$ (2,0) superconformal theory} in 6d
\cite{Witten:1998wy, Moore:2004jv, Belov:2006jd} 
-- the worldvolume theory of an M5-brane in M-theory.
The conjecture that these exotic QFTs can be consistently decoupled from gravity
underlies much recent progress in string theory (\eg~\cite{Gaiotto:2009we} and references thereto).
In particular, by compactification, its existence makes various deep 4d QFT dualities manifest.
This model reduces to the simple model without fermions or scalars when 
the partition function preserves enough supersymmetry \cite{Moore:2004jv}.
We conclude that the $(2,0)$ theory cannot be regularized preserving 
a sufficient amount of supersymmetry to allow this.
It is possible that further 
inquiry in this direction can provide guidance for better definitions of string theory.

\subsection{Would-be gauge anomalies and surface-only models}

Many surface-only obstructions,
including the Nielsen-Ninomiya examples, are directly related to {\it anomalies}
in the protecting symmetry: 
They would be gauge anomalies if we tried to gauge the protecting symmetry.
This property has been used to identify nontrivial SPT states
in \cite{2012Levin-Gu}.

We described an obstruction to regularizing 
a self-dual 2-form theory in $D=5+1$;
this, too, can be understood in terms of a known anomaly.
Just as for chiral CFTs 
(the chiral boson theory is a theory of a self-dual $1$-form field strength),
a {\it gravitational anomaly} was relevant here.
In 1+1 dimensions, the chiral central charge $c_L - c_R$, which determines the 
magnitude of the thermal Hall effect, measures an anomaly that 
obstructs coupling the theory to gravity.  
A similar gravitational anomaly 
arises for the 6d self-dual tensor theory \cite{AlvarezGaume:1983ig}.

Are all of surface-only obstructions to realizing a symmetric regulator
merely anomalies we would find should we try to gauge the 
symmetry?   
Thinking of 
anomalies as `symmetries broken by {\it any} regulator'
the answer would seem to be `yes' \cite{Wen:2013oza}.
It is true that the NN theorem can 
be interpreted as the statement that 
no chirally-symmetric regulator 
can produce the correct chiral anomaly in a 
background gauge field coupled to the vector current.

It is not clear, however, whether our present notion of `anomaly'
is general enough to make this the correct answer.
Here is some evidence that there can be obstructions more general than
obstructions to gauging symmetries:
\begin{enumerate}
\item A known example \cite{2012Levin-Gu}
is SPT states protected by time-reversal $\IZ_2^\CT$.  
It is not clear what it would mean to gauge an antiunitary symmetry 
(all numbers are real?), 
so it is not clear that obstructions to regularization arising from demanding $\IZ_2^\CT$ symmetry
can be thought of as would-be discrete gauge anomalies.

\item
From our discussion above, we can conclude that it is impossible to 
gauge electromagnetic duality in Maxwell theory.
Previous literature suggesting that it is impossible to gauge
EM duality includes \cite{Deser:2010it, Bunster:2011aw, Saa:2011wb}.
(In other models in other dimensions, it is possible to gauge analogs of EM duality \cite{MBXGW09}.)
We do not know how to describe this phenomenon in terms of a known 
$\IZ_2$ anomaly.
We note, however, that a model which resulted from identifying configurations related by the $S$ transformation
could not be CPT invariant,
since CPT in $4k$ spacetime dimensions relates helicity $\pm 1$ states.

We observe (following \cite{Maldacena:2001ss}) that 
the model we get on the surface is exactly in the 
form given by \cite{Schwarz:1993vs}.
We are forced to conclude that the manifestly EM-duality-invariant
(but not manifestly Lorentz-invariant)
model described in \cite{Schwarz:1993vs} cannot 
be regularized.
(Note that Lorentz invariance
is not the protecting symmetry: breaking Lorentz-invariance in the bulk
of the 5d CS theory
has no effect on the surface states.)

\item Finally, it would be very interesting to find 
obstructions to supersymmetry-preserving regulators.  Gauging supersymmetry leads to supergravity.  
Consistency of quantum supergravity 
is much harder to verify 
than the absence of quantum violations of the current.

\end{enumerate}

\vskip.2in
{\bf Acknowledgements:}
We would like to thank 
Maissam Barkeshli, Xiaoliang Qi, Congjun Wu, Cenke Xu
and especially T.~Senthil and Brian Swingle
for discussions.
We also thank Brian Swingle for collaboration on the related work \cite{BGSprogress},
and for comments on the manuscript.
JM is grateful to the Perimeter Institute for hosting the workshop
Entanglement and Emergence, II,
where part of this work was done.
This research was supported in part by
funds provided by the U.S. Department of Energy
(D.O.E.) under cooperative research agreements DE-FG0205ER41360
and DE-SC0009919,	
in part by the Alfred P. Sloan Foundation,
and in part by Perimeter Institute for Theoretical Physics. Research at Perimeter Institute is supported by the Government of Canada through Industry Canada and by the Province of Ontario through the Ministry of Economic Development $\&$ Innovation.


\appendix
\section{Appendix: A simple but non-unitary TFT}

Here we consider analogous gaussian models
where the fields are grassman-valued p-forms in 2p+1 dimensions:
$$ S = {K_{IJ} \over 4 \pi}  \int \gamma^I \wedge \dd \gamma^J .$$

Our motivation for thinking about this was that 
this reverses the behavior of the K matrices as a function of the parity of p,
relative to the case of bosonic forms.
That is: 
in 2+1 dimensions (odd p), the K-matrix is antisymmetric,
while in 4+1 dimensions (even p) the K-matrix is symmetric.

There is an analog in 1+1 dimensions which is called the $bc$ CFT,
an example of which arises in string theory.
There $b, c$ are grassmann fields with action $ \int b \bar \partial c$.
there one can consider the spin of $(b, c)$ to be $ (\lambda, 1-\lambda)$ respectively;
it is a CFT for any $\lambda$ with (chiral) central charge 
$c_\lambda = - 3 (2 \lambda - 1 )^2 + 1 $.  
$\lambda = \half$ is free complex fermions, $c_\half=1$.
The spin doesn't make any difference in terms of counting of degrees of freedom in 1+1d,
it just affects the form of the stress tensor.
If we take the spins to be $0, 1$ ($\lambda= 0$), so that we have a one-form and a zero-form 
we get $ c=-2$.  It is a non-unitary theory (the central charge is positive in a unitary CFT).

Such a system violates the usual connection between spin and statistics,
which is general for unitary relativistic quantum field theories
(indeed we will find that it is not unitary).
If we add a Maxwell-like term
${1\over m} \( \partial \gamma \)^2$ 
then surely this action will propagate ghosts.
But in the topological limit $ m \to 0$, nothing propagates,
and we might hope it is unitary, but has no relativistic UV completion.
As we will see by finding the space of states on a closed manifold, it is not.

If we study the quantization of this model on a closed manifold $\Sigma_{2p}$ we can expand
$$ \gamma^I = \sum_{ \alpha = 1}^{b^p(\Sigma_{2p}, \IZ) } \omega_\alpha \theta^{\alpha I} $$
and find
$$ S = \int \dd t {K_{IJ} \over 4 \pi} \CI_{\alpha\beta} \theta^{\alpha I} \dot \theta^{\beta J} ~,$$
with $\CI_{\alpha\beta} = \int_{\Sigma_{2p}} \omega_\alpha \wedge \omega_\beta $ the intersection form on $p$-cycles.
Quantizing this system leads to the canonical anticommutation relations
$$ \{ \theta^{\alpha I}, \theta^{\beta J } \} = 4 \pi \hbar  \( K^{-1} \)^{IJ} \( \CI^{-1} \)^{\alpha\beta} . $$
In any dimension $2p$ there are manifolds with indefinite signature of the intersection form on the middle homology.
If the RHS is of indefinite signature, the Hilbert space representing this algebra has states of negative norm.

\bibliographystyle{ssg}

\bibliography{nogo}

\begingroup\raggedright\begin{thebibliography}{10}

\bibitem{Nielsen:1980rz}
H.~B. Nielsen and M.~Ninomiya, ``{Absence of Neutrinos on a Lattice. 1. Proof
  by Homotopy Theory},'' {\em Nucl.Phys.} {\bf B185} (1981) 20.

\bibitem{Nielsen:1981xu}
H.~B. Nielsen and M.~Ninomiya, ``{Absence of Neutrinos on a Lattice. 2.
  Intuitive Topological Proof},'' {\em Nucl.Phys.} {\bf B193} (1981) 173.

\bibitem{Nielsen:1981hk}
H.~B. Nielsen and M.~Ninomiya, ``{No Go Theorem for Regularizing Chiral
  Fermions},'' {\em Phys.Lett.} {\bf B105} (1981) 219.

\bibitem{Friedan:1982nk}
D.~Friedan, ``{A PROOF OF THE NIELSEN-NINOMIYA THEOREM},'' {\em
  Commun.Math.Phys.} {\bf 85} (1982) 481--490.

\bibitem{wen04}
X.-G. Wen, {\em Quantum Field Theory of Many-Body Systems}.
\newblock Oxford Univ. Press, Oxford, 2004.

\bibitem{Wen:2012hm}
X.-G. Wen, ``{Topological order: from long-range entangled quantum matter to an
  unification of light and electrons},''
  \href{http://xxx.lanl.gov/abs/1210.1281}{{\tt 1210.1281}}.

\bibitem{LW0605}
M.~Levin and X.-G. Wen, ``Detecting Topological Order in a Ground State Wave
  Function,'' {\em Phys. Rev. Lett.} {\bf 96} (Mar, 2006) 110405.

\bibitem{KP0604}
A.~Kitaev and J.~Preskill, ``Topological Entanglement Entropy,'' {\em Phys.
  Rev. Lett.} {\bf 96} (Mar, 2006) 110404.

\bibitem{AshvinTarunTurner}
T.~{Grover}, A.~M. {Turner}, and A.~{Vishwanath}, ``{Entanglement entropy of
  gapped phases and topological order in three dimensions},'' {\em \prb} {\bf
  84} (Nov., 2011) 195120, \href{http://xxx.lanl.gov/abs/1108.4038}{{\tt
  1108.4038}}.

\bibitem{Turner:2013kp}
A.~M. Turner and A.~Vishwanath, ``{Beyond Band Insulators: Topology of
  Semi-metals and Interacting Phases},''
  \href{http://xxx.lanl.gov/abs/1301.0330}{{\tt 1301.0330}}.

\bibitem{K0602}
A.~Kitaev, ``Anyons in an exactly solved model and beyond,'' {\em Annals of
  Physics} {\bf 321} (2006), no.~1 2 -- 111. January Special Issue.

\bibitem{Kitaev-unpublished}
A.~Kitaev, ``unpublished,''.

\bibitem{Lu:2012dt}
Y.-M. Lu and A.~Vishwanath, ``{Theory and classification of interacting
  'integer' topological phases in two dimensions: A Chern-Simons approach},''
  {\em Phys.Rev.} {\bf B86} (2012) 125119,
  \href{http://xxx.lanl.gov/abs/1205.3156}{{\tt 1205.3156}}.

\bibitem{BGS-unpublished}
J.~Sau, B.~Swingle, and T.~Senthil, ``unpublished,''.

\bibitem{Kitaev:2009mg}
A.~Kitaev, ``{Periodic table for topological insulators and superconductors},''
  {\em AIP Conf.Proc.} {\bf 1134} (2009) 22--30,
  \href{http://xxx.lanl.gov/abs/0901.2686}{{\tt 0901.2686}}.

\bibitem{Ryu:2010zza}
S.~Ryu, A.~P. Schnyder, A.~Furusaki, and A.~W. Ludwig, ``{Topological
  insulators and superconductors: Tenfold way and dimensional hierarchy},''
  {\em New J.Phys.} {\bf 12} (2010) 065010.

\bibitem{Fidkowski-Kitaev}
L.~{Fidkowski} and A.~{Kitaev}, ``{Effects of interactions on the topological
  classification of free fermion systems},'' {\em \prb} {\bf 81} (Apr., 2010)
  134509, \href{http://xxx.lanl.gov/abs/0904.2197}{{\tt 0904.2197}}.

\bibitem{Senthil:2012tm}
T.~Senthil and M.~Levin, ``{Integer quantum Hall effect for bosons: A physical
  realization},'' {\em Phys.Rev.Lett.} {\bf 110} (2013) 046801,
  \href{http://xxx.lanl.gov/abs/1206.1604}{{\tt 1206.1604}}.

\bibitem{Vishwanath:2012tq}
A.~Vishwanath and T.~Senthil, ``{Physics of three dimensional bosonic
  topological insulators: Surface Deconfined Criticality and Quantized
  Magnetoelectric Effect},'' {\em Phys. Rev. X 3,} {\bf 011016} (2013)
  \href{http://xxx.lanl.gov/abs/1209.3058}{{\tt 1209.3058}}.

\bibitem{Wang:2013xqa}
C.~Wang and T.~Senthil, ``{Boson topological insulators: A window into highly
  entangled quantum phases},'' \href{http://xxx.lanl.gov/abs/1302.6234}{{\tt
  1302.6234}}.

\bibitem{Xu:2013bi}
C.~Xu and T.~Senthil, ``{Wave Functions of Bosonic Symmetry Protected
  Topological Phases},'' \href{http://xxx.lanl.gov/abs/1301.6172}{{\tt
  1301.6172}}.

\bibitem{ChenGuWen}
X.~{Chen}, Z.-C. {Gu}, Z.-X. {Liu}, and X.-G. {Wen}, ``{Symmetry protected
  topological orders and the group cohomology of their symmetry group},'' {\em
  ArXiv e-prints} (June, 2011) \href{http://xxx.lanl.gov/abs/1106.4772}{{\tt
  1106.4772}}.

\bibitem{Metlitski:2013uqa}
M.~A. Metlitski, C.~Kane, and M.~P.~A. Fisher, ``{Bosonic topological insulator
  in three dimensions and the statistical Witten effect},''
  \href{http://xxx.lanl.gov/abs/1302.6535}{{\tt 1302.6535}}.

\bibitem{Burnell:2013bka}
F.~Burnell, X.~Chen, L.~Fidkowski, and A.~Vishwanath, ``{Exactly Soluble Model
  of a 3D Symmetry Protected Topological Phase of Bosons with Surface
  Topological Order},'' \href{http://xxx.lanl.gov/abs/1302.7072}{{\tt
  1302.7072}}.

\bibitem{Wen:2013oza}
X.-G. Wen, ``{Classifying gauge anomalies through SPT orders and classifying
  anomalies through topological orders},''
  \href{http://xxx.lanl.gov/abs/1303.1803}{{\tt 1303.1803}}.

\bibitem{Callan:1984sa}
J.~Callan, Curtis~G. and J.~A. Harvey, ``{Anomalies and Fermion Zero Modes on
  Strings and Domain Walls},'' {\em Nucl.Phys.} {\bf B250} (1985) 427.

\bibitem{Kaplan:1992bt}
D.~B. Kaplan, ``{A Method for simulating chiral fermions on the lattice},''
  {\em Phys.Lett.} {\bf B288} (1992) 342--347,
  \href{http://xxx.lanl.gov/abs/hep-lat/9206013}{{\tt hep-lat/9206013}}.

\bibitem{Kaplan:2009yg}
D.~B. Kaplan, ``{Chiral Symmetry and Lattice Fermions},''
  \href{http://xxx.lanl.gov/abs/0912.2560}{{\tt 0912.2560}}.

\bibitem{Qi:2008ew}
X.-L. Qi, T.~Hughes, and S.-C. Zhang, ``{Topological Field Theory of
  Time-Reversal Invariant Insulators},'' {\em Phys.Rev.} {\bf B78} (2008)
  195424, \href{http://xxx.lanl.gov/abs/0802.3537}{{\tt 0802.3537}}.

\bibitem{Deser:1981wh}
S.~Deser, R.~Jackiw, and S.~Templeton, ``{Topologically Massive Gauge
  Theories},'' {\em Annals Phys.} {\bf 140} (1982) 372--411.

\bibitem{Schwarz:1978cn}
A.~S. Schwarz, ``{The Partition Function of Degenerate Quadratic Functional and
  Ray-Singer Invariants},'' {\em Lett.Math.Phys.} {\bf 2} (1978) 247--252.

\bibitem{Schwarz:1979ae}
A.~S. Schwarz, ``{The Partition Function of a Degenerate Functional},'' {\em
  Commun.Math.Phys.} {\bf 67} (1979) 1--16.

\bibitem{Witten:1988hf}
E.~Witten, ``{Quantum Field Theory and the Jones Polynomial},'' {\em
  Commun.Math.Phys.} {\bf 121} (1989) 351.

\bibitem{Elitzur:1989nr}
S.~Elitzur, G.~W. Moore, A.~Schwimmer, and N.~Seiberg, ``{Remarks on the
  Canonical Quantization of the Chern-Simons-Witten Theory},'' {\em Nucl.Phys.}
  {\bf B326} (1989) 108.

\bibitem{Horowitz:1989ng}
G.~T. Horowitz, ``{Exactly Soluble Diffeomorphism Invariant Theories},'' {\em
  Commun.Math.Phys.} {\bf 125} (1989) 417.

\bibitem{Blau:1989bq}
M.~Blau and G.~Thompson, ``{Topological Gauge Theories of Antisymmetric Tensor
  Fields},'' {\em Annals Phys.} {\bf 205} (1991) 130--172.

\bibitem{Horowitz:1989km}
G.~T. Horowitz and M.~Srednicki, ``{A QUANTUM FIELD THEORETIC DESCRIPTION OF
  LINKING NUMBERS AND THEIR GENERALIZATION},'' {\em Commun.Math.Phys.} {\bf
  130} (1990) 83.

\bibitem{Witten:1998wy}
E.~Witten, ``{AdS / CFT correspondence and topological field theory},'' {\em
  JHEP} {\bf 9812} (1998) 012,
  \href{http://xxx.lanl.gov/abs/hep-th/9812012}{{\tt hep-th/9812012}}.

\bibitem{Moore:2004jv}
G.~W. Moore, ``{Anomalies, Gauss laws, and Page charges in M-theory},'' {\em
  Comptes Rendus Physique} {\bf 6} (2005) 251--259,
  \href{http://xxx.lanl.gov/abs/hep-th/0409158}{{\tt hep-th/0409158}}.

\bibitem{Belov:2004ht}
D.~Belov and G.~W. Moore, ``{Conformal blocks for AdS(5) singletons},''
  \href{http://xxx.lanl.gov/abs/hep-th/0412167}{{\tt hep-th/0412167}}.

\bibitem{Hartnoll:2006zb}
S.~A. Hartnoll, ``{Anyonic strings and membranes in AdS space and dual
  Aharonov-Bohm effects},'' {\em Phys.Rev.Lett.} {\bf 98} (2007) 111601,
  \href{http://xxx.lanl.gov/abs/hep-th/0612159}{{\tt hep-th/0612159}}.

\bibitem{Mooretalk}
G.~Moore, ``{Minicourse of three lectures on Generalized Abelian Gauge
  Theories, Self-Duality, and Differential Cohomology, at the Simons Center
  Workshop on Differential Cohomology, Simons Center for Geometry and Physics,
  Stonybrook},''.

\bibitem{Freed:2006yc}
D.~S. Freed, G.~W. Moore, and G.~Segal, ``{Heisenberg Groups and Noncommutative
  Fluxes},'' {\em Annals Phys.} {\bf 322} (2007) 236--285,
  \href{http://xxx.lanl.gov/abs/hep-th/0605200}{{\tt hep-th/0605200}}.

\bibitem{Gross:1998gk}
D.~J. Gross and H.~Ooguri, ``{Aspects of large N gauge theory dynamics as seen
  by string theory},'' {\em Phys.Rev.} {\bf D58} (1998) 106002,
  \href{http://xxx.lanl.gov/abs/hep-th/9805129}{{\tt hep-th/9805129}}.

\bibitem{Witten:1998xy}
E.~Witten, ``{Baryons and branes in anti-de Sitter space},'' {\em JHEP} {\bf
  9807} (1998) 006, \href{http://xxx.lanl.gov/abs/hep-th/9805112}{{\tt
  hep-th/9805112}}.

\bibitem{Shatashvili-unpublished}
S.~Shatashvili, ``unpublished,''.

\bibitem{Witten:1996md}
E.~Witten, ``{On flux quantization in M theory and the effective action},''
  {\em J.Geom.Phys.} {\bf 22} (1997) 1--13,
  \href{http://xxx.lanl.gov/abs/hep-th/9609122}{{\tt hep-th/9609122}}.

\bibitem{Maldacena:2001ss}
J.~M. Maldacena, G.~W. Moore, and N.~Seiberg, ``{D-brane charges in five-brane
  backgrounds, Appendix A},'' {\em JHEP} {\bf 0110} (2001) 005,
  \href{http://xxx.lanl.gov/abs/hep-th/0108152}{{\tt hep-th/0108152}}.

\bibitem{Belov:2005ze}
D.~Belov and G.~W. Moore, ``{Classification of Abelian spin Chern-Simons
  theories},'' \href{http://xxx.lanl.gov/abs/hep-th/0505235}{{\tt
  hep-th/0505235}}.

\bibitem{Belov:2006jd}
D.~Belov and G.~W. Moore, ``{Holographic Action for the Self-Dual Field},''
  \href{http://xxx.lanl.gov/abs/hep-th/0605038}{{\tt hep-th/0605038}}.

\bibitem{2004AnPhy.313..497H}
T.~H. {Hansson}, V.~{Oganesyan}, and S.~L. {Sondhi}, ``{Superconductors are
  topologically ordered},'' {\em Annals of Physics} {\bf 313} (Oct., 2004)
  497--538, \href{http://xxx.lanl.gov/abs/arXiv:cond-mat/0404327}{{\tt
  arXiv:cond-mat/0404327}}.

\bibitem{DonaldsonBook}
S.~K. Donaldson and P.~B. Kronheimer, {\em The Geometry of Four-Manifolds}.
\newblock Oxford Univ. Press, Oxford, 1990.

\bibitem{BGSprogress}
S.~Kravec, J.~McGreevy, and B.~Swingle, ``work in progress,''.

\bibitem{Kane:1994vb}
C.~Kane, M.~P. Fisher, and J.~Polchinski, ``{Randomness at the edge: Theory of
  quantum Hall transport at filling nu = 2/3},'' {\em Phys.Rev.Lett.} {\bf 72}
  (1994) 4129--4132.

\bibitem{Kane:1994jd}
C.~Kane and M.~P. Fisher, ``{Edge state transport},''.

\bibitem{Schwarz:1993vs}
J.~H. Schwarz and A.~Sen, ``{Duality symmetric actions},'' {\em Nucl.Phys.}
  {\bf B411} (1994) 35--63, \href{http://xxx.lanl.gov/abs/hep-th/9304154}{{\tt
  hep-th/9304154}}.

\bibitem{LW0510}
M.~A. Levin and X.-G. Wen, ``String-net condensation: A physical mechanism for
  topological phases,'' {\em Phys. Rev. B} {\bf 71} (Jan, 2005) 045110.

\bibitem{Gaiotto:2009we}
D.~Gaiotto, ``{N=2 dualities},'' {\em JHEP} {\bf 1208} (2012) 034,
  \href{http://xxx.lanl.gov/abs/0904.2715}{{\tt 0904.2715}}.

\bibitem{2012Levin-Gu}
M.~{Levin} and Z.-C. {Gu}, ``{Braiding statistics approach to
  symmetry-protected topological phases},'' {\em \prb} {\bf 86} (Sept., 2012)
  115109, \href{http://xxx.lanl.gov/abs/1202.3120}{{\tt 1202.3120}}.

\bibitem{AlvarezGaume:1983ig}
L.~Alvarez-Gaume and E.~Witten, ``{Gravitational Anomalies},'' {\em Nucl.
  Phys.} {\bf B234} (1984) 269.

\bibitem{Deser:2010it}
S.~Deser, ``{No local Maxwell duality invariance},'' {\em Class.Quant.Grav.}
  {\bf 28} (2011) 085009, \href{http://xxx.lanl.gov/abs/1012.5109}{{\tt
  1012.5109}}.

\bibitem{Bunster:2011aw}
C.~Bunster and M.~Henneaux, ``{Sp(2n,R) electric-magnetic duality as off-shell
  symmetry of interacting electromagnetic and scalar fields},'' {\em PoS} {\bf
  HRMS2010} (2010) 028, \href{http://xxx.lanl.gov/abs/1101.6064}{{\tt
  1101.6064}}.

\bibitem{Saa:2011wb}
A.~Saa, ``{Local electromagnetic duality and gauge invariance},'' {\em
  Class.Quant.Grav.} {\bf 28} (2011) 127002,
  \href{http://xxx.lanl.gov/abs/1101.3927}{{\tt 1101.3927}}.

\bibitem{MBXGW09}
M.~{Barkeshli} and X.-G. {Wen}, ``{Effective field theory and projective
  construction for Z$_{k}$ parafermion fractional quantum Hall states},'' {\em
  \prb} {\bf 81} (Apr., 2010) 155302,
  \href{http://xxx.lanl.gov/abs/0910.2483}{{\tt 0910.2483}}.

\end{thebibliography}\endgroup

\end{document}